\documentclass[preprint,authoryear,12pt]{elsarticle}

\usepackage{graphicx}
\biboptions{longnamesfirst,comma}
\journal{Planetary \& Space Science}

\begin{document} 
\begin{frontmatter}

\title{Tidal interactions - crude body model in dynamical investigations}

\author[rg]{Ryszard Gabryszewski}
\ead{r.gabryszewski@cbk.waw.pl}
\address[rg]{Space Research Centre \\ Polish Academy of Sciences \\ ul. Bartycka 18 A \\ 00-716 Warszawa, Poland}


\begin{abstract}
The paper presents results of investigations of small bodies dynamics in a vicinity of giant planets. We used the most simple body model: gravitationally bounded, rotating contact binary affected by the tidal force acting from a planet. Spin variations of such binaries were extensively studied during planetary close encounters. Two main types of dynamical behaviour were observed: (i) huge but interim fluctuations of the angular velocity and (ii) permanent changes of a rotation during a close approach. The first type is observed mainly for fast rotators, while the second one was encountered in a population of slowly spinning objects with periods longer than 12 hours. Conclusions on usability of such crude physical body models in dynamical investigations and a comparison to previous results were  attached. The results allow us to formulate a thesis explaining the phenomenon of creation of the extremely slow rotators and an observational excess of such type of objects. 
\end{abstract}
\end{frontmatter}

\section{Introduction}

The phenomenon of tidal interactions was widely discussed in different models over last 160 years. Last decade of the XX century brought important papers which extensively explored the problem of tidal effects in a vicinity of planets \citep{key-cha,key-brml}. Studies of small bodies dynamics in such region are usually made using the most complex physical body models which are able to give output in reasonable time scales. Test bodies are most often conglomerates consisting of hundreds or even thousands small and equal spheres interacting gravitationally with each other \citep{key-19,key-20}, also often affected by the forces of viscosity. Such multiparameter models describe alterations of the physical structure very well but studies of dynamical processes during planetary approaches do not always need complex models. Reduction of model parameters is perceived as a loss of a part of information but crude models also have some advantages. They allow to get general information on physical phenomenons much faster than complex ones and permit to find dependencies between model parameters and type of a body's behaviour with much smaller numerical effort. Another reason is the ease of implementation of features which are hard to apply in complex structures, e.g. an inhomogeneity defined as difference in densities and/or diameters of the body's components. 

The main aim of this paper is to investigate dynamics of a contact binary during close approaches to a giant planet. A contact binary can be treated as a toy model of a rubble-pile body and easily allows to include an inhomogeneity of its components. In contrary to recent papers we investigated tidal dynamics of small bodies with diameters similar to size of Centaurs and cometary densities. We also tried to find out how simplification of a physical body model affected information on dynamical evolution in a reference to previous works.

\section{Model and calculations}

The model of a small body consists of two spherical components held together by autogravity
forces. Both fragments contact each other and revolve their common mass centre. This binary system approaches a planet at a small but non-collisional minimum distance. Dynamics of the binary is considered in a N-body problem with Sun and a giant planet on an elliptic orbit. All the bodies in the system move in only one plane - the orbital motion of the binary can be prograde and retrograde while the motion of a planet can only be prograde. The binary rotates in the plane of orbital motion of the planet. This means that the torque direction and angular momentum vector are always perpendicular to that plane. 

\begin{figure}
\centerline{\includegraphics{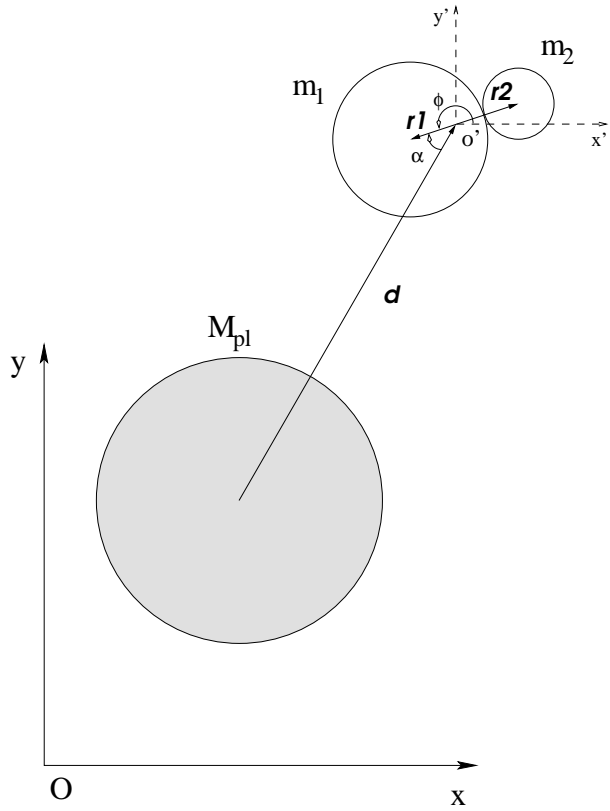}}
\caption{Model of a test body consists of masses $m_{1}$ and $m_{2}$ approaching a planet $M_{pl}$. Orbital motion of all the bodies is considered in the mass centre reference frame. Rotation of the binary is investigated in the frame related to its mass centre: O'. Vector ${\bf d}$ determines distance and direction between $M_{pl}$ and the mass centre of the binary. Phase angle $\phi$ defines position of the binary in non-rotating Cartesian reference frame O'x'y' while $\alpha$ is an orientation angle of the binary in respect to ${\bf d}$ vector.}
\label{rys1}
\end{figure}

\subsection{Orbital and rotational motion}

Two types of reference systems were used: orbital motion was considered in the mass centre reference frame of all the bodies while rotation of the binary was investigated in a frame related to its own mass centre, see Fig. \ref{rys1}. The equation of orbital motion can be expressed as follows:

\begin{eqnarray}
\ddot{\vec{r}_{i}} = -k^{2} \sum_{j=1}^{N} m_{j} \frac{\vec{r}_{ij}}{r_{ij}^{3}}
\label{eq-1}
\end{eqnarray}

\noindent
where i=1,..,N and N is the number of bodies in the system, $ \vec{r}_{ij} = \vec{r}_{i} - \vec{r}_{j}$, $i \neq j$, $k$ is the gaussian gravitational constant, $m_{j}$ is the mass of $j$th body, and $\vec{r}_{i}$ and $\vec{r}_{j}$ are position vectors of $i$th and $j$th body respectively. The equation was solved numerically using the recurrent power series (RPS) method \citep{key-9}. We extended the original 2-body algorithm to solve N-body problem. The auxiliary variables were defined as follows:

\begin{eqnarray}
\vec{p}_{i} = {\dot{\vec{r}_{i}}},\quad   s_{ij} = r^{2}_{ij},\quad
u_{ij} = r_{ij}s_{ij},\quad \vec{q}_{ij} = \frac{\vec{r}_{ij}}{u_{ij}}
\label{eq-2}
\end{eqnarray}

\noindent
where $\dot{\vec{r}_{i}}$ is the vector of velocity of the $i$th body. If we put them to the equation \ref{eq-1}, we get:

\begin{eqnarray}
\dot{\vec{p}_{i}} = -k^{2} \sum_{j=1}^{N} m_{j} \vec{q}_{ij}
\label{eq-3}
\end{eqnarray}

\noindent
If we designate the derivatives:

\begin{eqnarray}
\dot{s}_{ij} = 2\vec{r}_{ij}\cdot\dot{\vec{r}}_{ij} \quad \Rightarrow \quad r_{ij}\dot{r}_{ij} = \frac{1}{2}\dot{s}_{ij}
\label{eq-4}
\end{eqnarray}

\noindent
we can determine the recurrent dependencies (written in the order of execution):

\begin{eqnarray}
\vec{r}_{i,n+1} & = & \frac{1}{n} \hspace{1mm} \vec{p}_{i,n}, \nonumber \\
s_{ij,n+1} & = & \frac{2}{n} \sum_{l=1}^{n} \vec{r}_{ij,l} \hspace{1mm} \vec{p}_{ij,n-l+1}, \nonumber \\
r_{ij,n+1} & = & \frac{s_{ij,n+1}}{2r_{ij,1}} - \frac{1}{nr_{ij,1}} \sum_{l=2}^{n} (n-l+1) \hspace{1mm} r_{ij,l} \hspace{1mm} r_{ij,n-l+2}, \nonumber \\
u_{ij,n+1} & = & \sum_{l=1}^{n+1} s_{ij,l} \hspace{1mm} r_{ij,n-l+2}, \\
\vec{q}_{ij,n+1} & = & \frac{\vec{r}_{ij,n+1}}{u_{ij,1}} - \frac{1}{u_{ij,1}} \sum_{l=2}^{n+1} u_{ij,l} \hspace{1mm} \vec{q}_{ij,n-l+2}, \nonumber \\
\vec{p}_{i,n+1} & = & - \frac{1}{n} \hspace{1mm} k^{2} \sum_{j=1}^{n} m_{j} \hspace{1mm} \vec{q}_{ij,n} \nonumber
\label{eq-5}
\end{eqnarray}

\noindent
where n is the coefficient number of the recurrent power series. The procedure of integration was enriched by the function for an automatic optimum time step calculation \citep{key-10}. The value of the time step is determined using last two coefficients of the recurrent power series:

\begin{eqnarray}
A_{n-1} & = & |x_{n-1}| + |y_{n-1}| + |z_{n-1}| \nonumber \\
A_{n} & = & |x_{n}| + |y_{n}| + |z_{n}|
\label{eq-6}
\end{eqnarray}

\noindent
where $x$, $y$ and $z$ are rectangular coordinates of a body (z coordinate is zero in the case considered in this paper). Then time step can be defined as:

\begin{eqnarray}
h & = & \left[  \frac{\varepsilon_{loc}}{A_{n-1}+A_{n}h_{0}(1+\frac{A_{n}}{A_{n-1}}h_{0})} \right]^{\frac{1}{n-1}}
\label{eq:6}
\end{eqnarray}

\noindent
where:

\begin{eqnarray*}
h_{0} & = & \left( \frac{\varepsilon_{loc}}{A_{n-1}} \right)^{\frac{1}{n-1}}
\end{eqnarray*}

\noindent
and $\varepsilon_{loc}$ is the wanted accuracy of numerical solution. The formula (\ref{eq:6}) allows us to determine the time step for every body in the integration process, we should choose the smallest value. Positions and velocities of the $i$th body can be obtained from the equations below:

\begin{eqnarray}
\vec{r}_{i} & = & \sum_{j=1}^{n} r_{j} \hspace{1mm} h^{j-1}, \\
\dot{\vec{r}_{i}} & = & \sum_{j=1}^{n} j r_{j+1} \hspace{1mm} h^{j-1} \nonumber
\label{eq-7}
\end{eqnarray}

Variations of a binary's angular velocity were investigated using Euler equations. In the planar problem where we have only one component of the angular velocity, we can simplify these equations to the form:

\begin{equation}
D = I \dot{\omega}
\label{eq-7a}
\end{equation}

\noindent
where $I$ is the total moment of inertia of a binary, $D$ is its torque and $\dot{\omega}$ is the angular acceleration of the body. The moment of inertia is defined as a sum of moments of both fragments:

\begin{equation}
I = \sum_{l=1}^{2} (\frac{2}{5}m_{l}d_{l}^{2} + m_{l}r_{l}^{2})
\end{equation}

\noindent
where $m_{l}$ are the masses of conglomerate's components, $d_{l}$ are the radii of $m_{l}$, and $r_{l}$ are the lever arms. The total torque is defined as follows:

\begin{equation}
\vec{D} = \sum_{l=1}^{2} \vec{r}_{l} \times \vec{F}_{Mm_{l}}
\label{eqrot-3}
\end{equation}

\noindent
where $F_{Mm_{l}}$ are forces acting from a planet $M$ to masses $m_{l}$ respectively. 

The equation (\ref{eq-7a}) was solved numerically by the Runge-Kutta method of the 5th order (RK5) with automatic time step calculations using standard algorithms published in the Numerical Recipes \citep{key-11}. Minor modifications were made to get a double precision output. Two different time scales of orbital and rotational motion were connected by invoking the RK5 integrator on every substep of the RPS method.

\subsection{Breakup of a contact binary and reconnection of the fragments}

We defined the tidal disruption as a separation of fragments due to tidal force. This separation is accelerated by a centrifugal force caused by the binary's rotation. Fragments start to be treated as separate bodies when the difference of accelerations acting from the planet on binary's components - increased by the centrifugal acceleration - exceeded the mutual acceleration in the $m_{1}$$m_{2}$ system:

\begin{equation}
a<|a_{1}-a_{2}|+a_{cf}
\label{disrpt}
\end{equation}






\noindent
where $a$ is a gravitational acceleration of $m_{1}$ respectively to $m_{2}$,
$a_{cf}$ is an acceleration due to centrifugal force and
$a_{1}$ and $a_{2}$ are accelerations of $m_{1}$ and $m_{2}$
respectively caused by gravitational force acting from the planet.

When fragments are separated their angular momentum do not change in time because of their spherical shape. The fragments can be merged again by mutual collisions. We assumed collisions to be pure inelastic in our model, this means that every mutual collision of fragments creates back a contact binary. This body can be disrupted in the next step of an integration process if the condition (\ref{disrpt}) is still fulfilled. These assumptions allowed to calculate new angular momentum and angular velocity of the body in moments of its creation and disruption.

\subsection{Initial conditions}

The investigated systems consisted initially of 3 bodies: Sun, a planet (Jupiter and Neptune in the calculations) and a contact binary. We generated 1400 sets of initial conditions. Starting perihelia of contact binaries were randomly spread in a distance to 0.003 AU from the perihelion of a planet, and the aphelia were defined between 9 and 45 AU: the bodies (a planet and a contact binary) started the dynamical evolution close to their perihelia. Their mean anomalies had initial values defined between 357 and 3 degrees (which defined initial distances between planet and the centre of the mass of the binary between 0.1 and 0.9 AU). Values of orbital inclinations of binaries were set to 0 and 180 degrees - prograde and retrograde motion was investigated. Masses of contact binaries were parametrized by densities and diameters of the fragments. The ranges of fragments diameters were defined between 50 and 100 km which interacts very well with diameters of the largest Centaurs: 2060 Chiron and 5145 Pholus \citep{key-cent}. The ranges of densities were chosen very wide, from low cometary 0.2 g/cm$^{3}$ \citep{key-2} to Centaur values of 1.5 g/cm$^{3}$ \citep{key-grundy}.

Most of previous papers studied the problem of tidal encounter in a 2-body problem where the small body moved on parabolic or hyperbolic orbit with a planet in its focus. Initial conditions were determined by small body's velocity "at infinity" v$_\infty$ and by periapse distance q. In our initial conditions v$_{\infty}$ varied in a range between 3 and 4 km/s for prograde motion and 30 to 34 km/s for retrograde motion. The minimum distance between a planet and a binary could only be known a'posteriori. 

We investigated variations of orbital elements of all the bodies in the system, angular velocity and geometry of the contact binary together with variations of mutual distance between its fragments after the breakup. Equations of motion were integrated up to 6 subsequent planetary encounters. 

\section{Results}

\begin{figure}
\centerline{\includegraphics{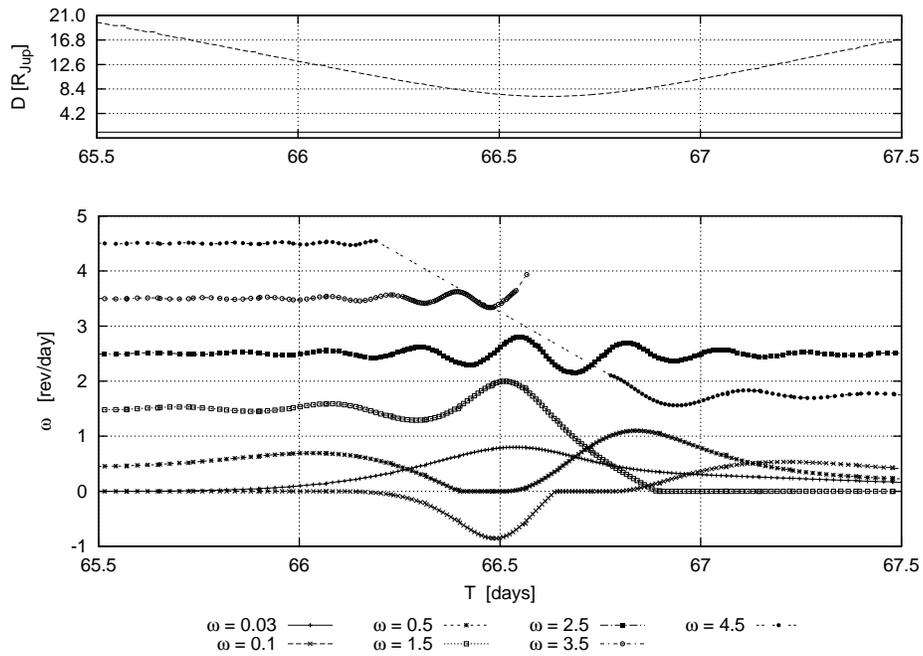}}
\caption{Variations of angular velocity of a contact binary in a function of time for different starting values. Upper figure shows the distance between Jupiter and the small body.  Initial distance between the planet and the body is 0.197 AU, while the minimum distance of the approach is 0.0034 AU ($\sim$7 planetary radii). Solid flat line on the upper chart denotes the planetary radius. The curves are interrupted when the binary breaks up. After the break-up event fragments are able to adjoin (case of $\omega$=4.5 rev/day) creating a new contact binary with much lower angular velocity.}
\label{f1}
\end{figure}

The simulations allowed to distinguish two frequently observed types of behaviour: oscillations
around initial value in case of fast (above 4 rev/day) and moderate ($\sim$2.5 rev/day) rotators and permanent changes of the angular velocity of slow rotators (see Fig. \ref{f1}). Small bodies with starting angular velocity of 0.1 rev/day or lower can even change their spin directions. The centrifugal force can be an important factor which helps tides to disrupt fast rotators, see the evolution of bodies with initial spin of 3.5 and 4.5 rev/day (the self disruption due to sole centrifugal force is about 5.2 rev/day in presented case). The reconnection of fragments is stable if the spin velocity of newly created binary is much lower than the initial one (see the case of $\omega$=4.5 rev/day). Distinct variations of a binary's spin were observed in distances up to 10 planetary radii for every investigated giant planet. We noticed that the influence of tidal force was nearly insignificant in a distance above 25 planetary radii (standard values of planetary equatorial radii were accepted \citep{key-std}).

The difference in spin variations between fast and slow rotators are the derivative of the torque $D$ changes.
The total torque changes its sign 4 times while an aspherical body revolves around its spin axis. This means the resultant torque is usually small for fast rotating objects but for slow rotators the change of torque is
able to achieve much larger values. Modifications of angular velocities are done in
a quite short time: 12 to 24 hours. When a distance to a planet is greater than 0.01 AU, torque makes only insignificant modifications to angular velocity - changes seem to be random and do not accumulate during subsequent close-ups. 

\begin{figure}
\centerline{\includegraphics{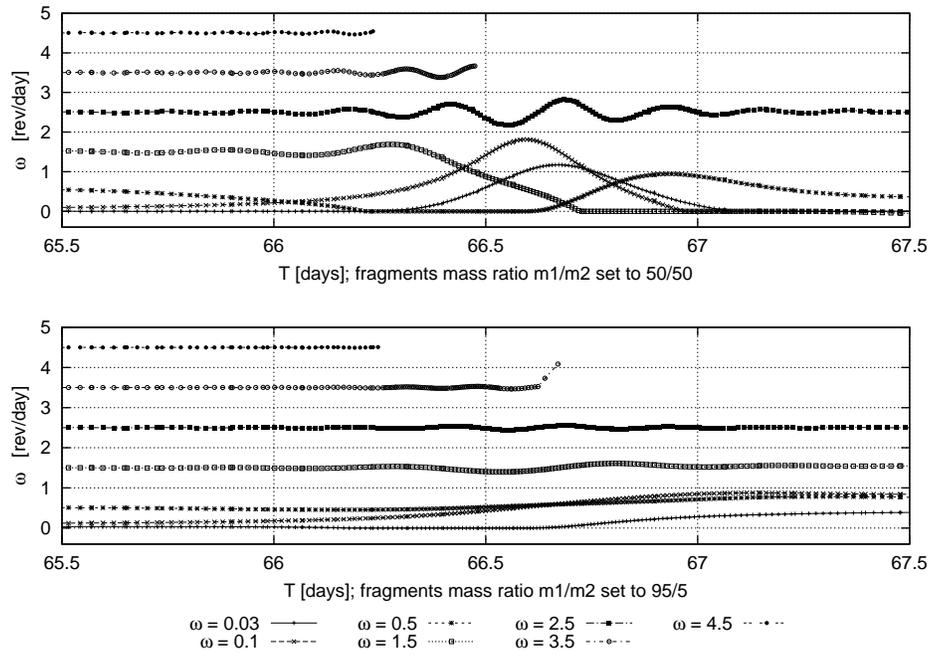}}
\caption{Variations of contact binary's angular velocity in a function of time during close encounter to Jupiter. 
Upper figure shows the diagram for identical masses of fragments, lower figure is for mass ratio of 95/5 (assuming the same sum of masses and the same fragments densities in both cases). The curves are interrupted when the binary breaks up.}
\label{run3Bfig}
\end{figure}

The contact binary body model allowed to incorporate an inhomogeneity in a very simple way. We made lots of simulations with fragments of different mass ratio (assuming the same sum of the masses and fragments densities). The general observed dependence between evolution of angular velocity and mass ratio of fragments is: the larger disproportion between masses, the lower amplitude of velocity variations, see Fig.~\ref{run3Bfig}. Distinct difference in the amplitude of spin velocities was observed if masses of fragments differed more than 15 \% 
. Such dependence is an effect of the body's shape. If the fragments have comparable masses and sizes, the shape of the body is the most elongated, contact binary formed by such fragments can be treated as an approximation of the most possible aspherical body created by 2 spheres. The influence of torque $D$ is the largest in such a case and gives us in effect largest possible variations of spin velocity and distance of disruption.

\begin{figure}
\centerline{\includegraphics{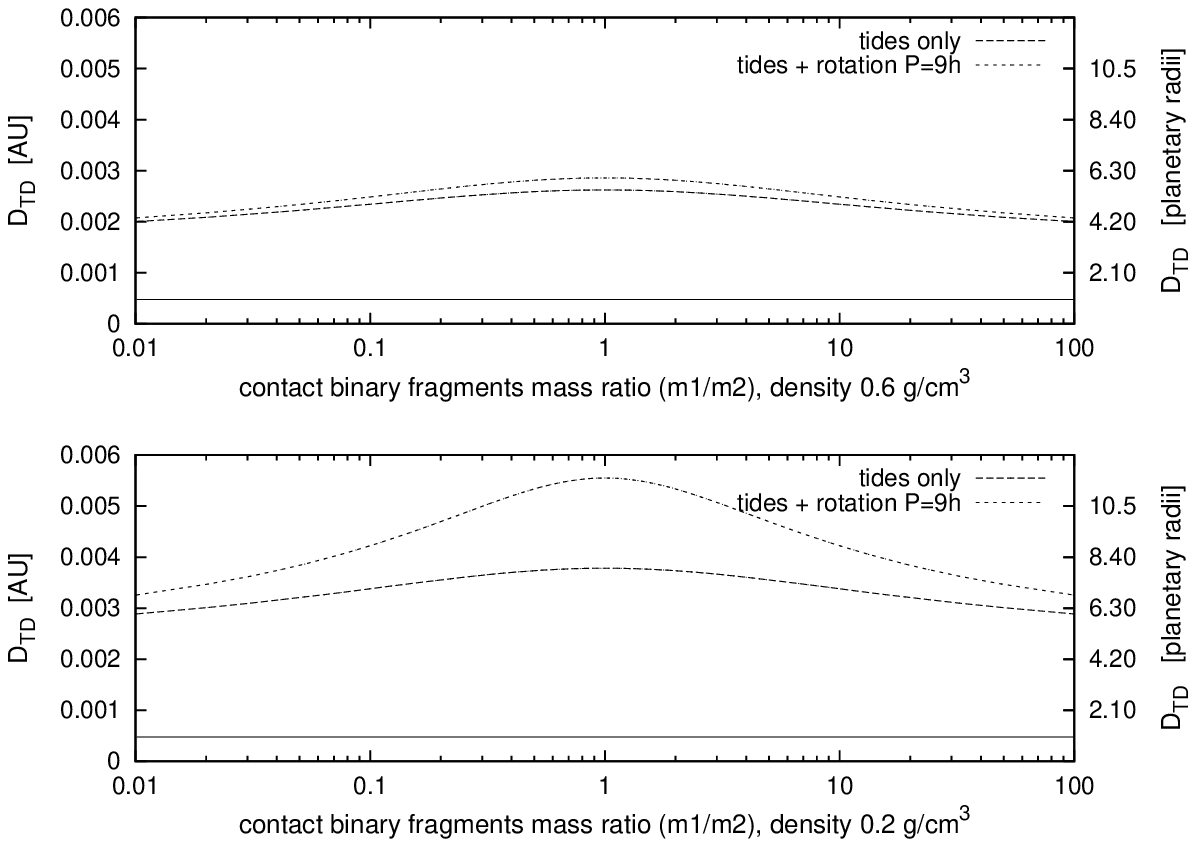}}
\caption{Maximum distance of tidal disruption in a function of fragments mass ratio for non-rotating contact binary (when sole tidal interactions were taken in account, tides only description on charts) and rotating with angular velocity of 2.67 rev/day (tides + rotation P=9h). Binary's densities were set to 0.6 g/cm$^{3}$ (upper chart) and 0.2 g/cm$^{3}$ (lower chart). Solid flat line on the figures denotes the planetary radius of Jupiter.}
\label{figtilt4}
\end{figure}

Fig.~\ref{figtilt4} shows the maximum distances of tidal disruption for contact binaries with densities: 0.2 g/cm$^{3}$ (lower graph) and 0.6 g/cm$^{3}$ (upper graph). Both graphs include curves showing the maximum distance of disruption for non-rotating binary (where tidal interactions only were taken into account) and rotating with spin velocity of 2.67 rev/day (tidal interactions + rotation P=9h). The figure shows that distance of disruption depends mainly on fragments mass ratio and their densities. Fast rotation can significantly increase this distance only when the binary has a very low density and consists of fragments of similar masses.

\begin{figure}
\centerline{\includegraphics{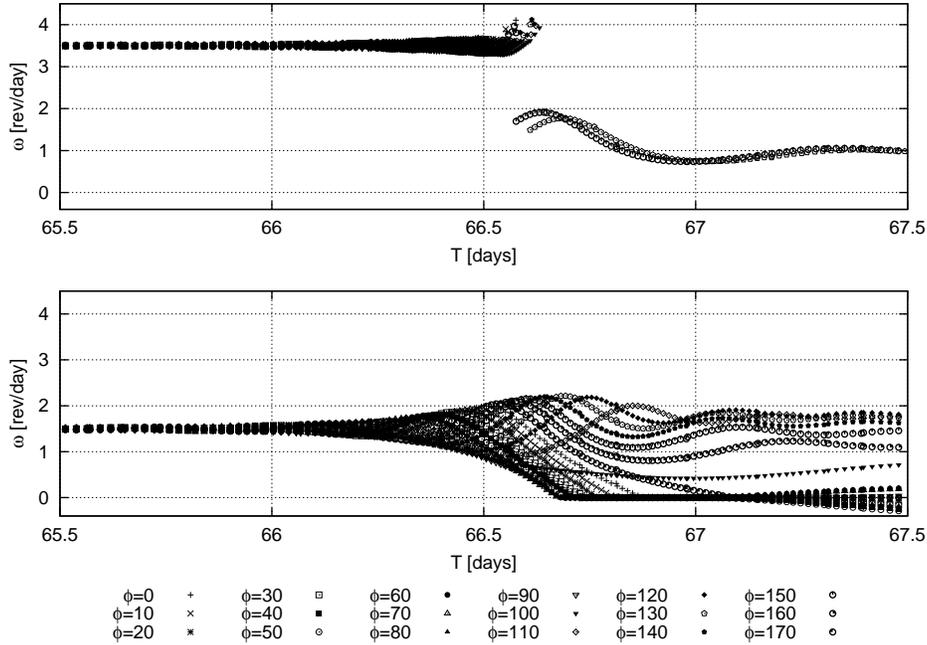}}
\caption{Variations of a contact binary's rotation as a function of time. Starting
value of angular velocity is set to 3.5 rev/day on upper graph and 1.5 rev/day on lower graph. Fast rotators can break up close to a minimum distance of planetary approach, the time of disruption depends on the phase angle $\phi$. After the break-up event fragments are able to adjoin (cases of $\phi$=130 and 150 deg) creating a new contact binary which rotates with much lower angular velocity. On lower graph most of slow rotators tend to lower their angular velocities after the close approach. Binaries with starting $\phi$ between 10 and 50 deg changed their spin directions permanently.}
\label{f5}
\end{figure}

Fig.~\ref{f5} shows variations of angular velocity in a function of a phase angle $\phi$. The computations were made for different starting values of angular velocity but the figure shows only the cases of $\omega$ = 1.5 and 3.5 rev/day. Two main types of the behaviour were observed. Below some threshold value  (2.5 rev/day in the presented case) rotation is slowing down for most values of the $\phi$ angle, the change of the spin direction is also possible (see lower graph). Although for some specific range of the $\phi$ angles - 100 to 150 degrees - angular velocity could even be increased up to $\sim$30 per cent. That geometry-dependent effect is related to an elongated shape of a contact binary and the changes of the torque. The upper graph shows that fast rotators are able to increase their angular velocity regardless of a starting value of $\phi$. If condition (\ref{disrpt}) is fulfilled, the binary disrupts and fragments start to evolve separately. Time of disintegration depends on the phase angle. In some cases fragments can adjoin and create a new binary rotating with an angular velocity much lower than the initial one. 

Tidal force can be responsible for observed excess of slow rotators \citep{key-13} but there are also other processes as like non-tidal breakups, non-tidal splitting \citep{key-harris} and YORP effect \citep{key-bottkevar} which can also create slow rotating bodies. Our results show that tidal interactions can be an effective process of slowing down low density objects in a vicinity of giant planets - in distances close to 7 - 8 planetary radii and smaller. In contrary to Weidenschilling's tidal despinning (1989), tidal interactions with giant planets can be treated as the mechanism able to produce objects with extremely long rotational periods, even comparable to the one of 288 Glauke - over 50 days \citep{key-mich}. 

Fast rotators are far less susceptible for permanent changes of angular velocity but such type of dynamics was also observed. The increase of the spin velocity leads to tidal disruption in most studied cases.

The results presented on figs.~\ref{f1},~\ref{run3Bfig} and \ref{f5} concern contact binaries with very low densities of 0.2 g/cm${^3}$. But general dependencies for the bodies with densities of 0.6 and 1.5 g/cm$^{3}$ are very similar. The differences concern the distances of disruption which are 1.5 to 2 times smaller respectively. Decrease of the breakup distance strongly affects variations of angular velocity - much higher amplitudes are observed. The centrifugal force hardly increases the disruption distance for bodies with densities of 0.6 g/cm${^3}$ and greater (see Fig. \ref{figtilt4}).

About 33\% of tested binaries were disrupted during the planetary approaches. In 72\% of those cases, one of the disrupted fragments hit the planet while another passed it by on a highly eccentric orbit. Both fragments survived the planetary close-ups in about 27\% cases. We also found very peculiar dynamics: double systems were created in about 1\% of cases. The separation of the fragments varied from tens of meters to about 180 km. Orbits of such double system had very elliptic shape: their eccentricities varied between 0.1 to 0.6 with higher values mainly for compact system. Such system were very unstable, they were disintegrated in a few to several days. The rest of double systems were stable in a timescale of computations but their mass centres moved on hyperbolic orbits which means they will leave the planetary system. Those peculiar double systems were created by collisions which were defined to be pure inelastic in our model.

\section{Conclusions}

The conclusions can be summarized as follows:

\begin{itemize}
\item We can distinguish two main types of behaviour of contact binaries angular velocities approaching giant planets:

(i) large but temporal amplitude of variations during a close approach without significant modification of the initial value after encounter, 
(ii) permanent change of a spin velocity of slow rotators with periods longer than 12 hours. 

\item Modifications of angular velocity are made in 12 to 24 hours during the close up. 

\item Binaries with angular velocities of 0.1 rev/day or lower can change the spin direction.

\item Significant variations of angular velocities can be observed up to $\sim$10 planetary radii due to tidal interactions, variations above this limit are random and do not accumulate. Tidal force at distances of 25 planetary radii or greater have a very weak influence on angular velocity.

\item Distance of tidal disruption depends on fragments densities and their mass ratio. Centrifugal force is able to extend this distance only in case of very low densities.

\item Extremely long period of rotation can be an effect of deep encounter between an aspherical small body and a giant planet. Spin periods of slow rotators can even decrease to months ($\omega$ $<$ 0.01 rev/day). 
\end{itemize}

\subsection{Discussion of results}

In Sharma's paper (2006), Figs. 8 and 10 show variations of angular velocity in a function of time during planetary close-ups. Although the data are in non-dimensional units we can simple convert them to our units using the information placed below the diagrams. The body which is initially a fast rotator ($\omega$=4.4 rev/day) is able to decrease its spin velocity to a value of $\omega$=1.84 rev/day due to tidal interactions. Initial moderate or slow rotators are able both to substantially increase their periods to $\omega$=4.52 rev/day or slow it down to $\omega$=0.6 rev/day. These levels of variations are in a very good agreement with the results presented in this work. We confirm that fast rotators can also essentially decrease their spin velocity but this type of evolution is observed seldom in our calculations.

The time of spin rate changes is about 7 hours at Sharma's results comparing to 12 - 24 hours in our outcomes. This difference can be explained by the minimum planetary distance during the approach: 1.8 planetary radii at Sharma's paper vs 7 planetary radii in our outcomes (v$_\infty$ was about 3 km/s both in Sharma's and our calculations), and also by a difference in the mass of an approached planet. 

Unfortunately, we cannot do similar comparison to most of previous papers \citep{key-17,key-18,key-20}. The authors described mainly the results of disruptions, i.e. the number of created fragments and their general dynamical properties, the works do not include numbers for angular valocity variations during planetary approach. But general dependencies in this work are fully consistent with the outcomes of the papers of Walsh \& Richardson (2006), related to test bodies with flattened ellipsoidal shape. 

It should be noticed that Solar System model used in Sharma's and Walsh's papers were completely different comparing to assumptions made in this work: their calculations were made in a 2-body problem where small body approached a planet on a hyperbolic orbit. But their dynamical conclusions are nearly the same as ours. The presented outcome is also consistent with Sekanina's works \citep{key-26,key-27,key-12}. In those papers they considered the importance of centrifugal force as a factor responsible for disruption of low density (0.2 g/cm$^{3}$) comets during close encounters with a planet. Their results also stay in good agreement with our outcomes. 

The results presented in this work show that detail investigations of objects dynamics can be driven using quite crude physical body models if important dynamical processes are included.

\section*{Acknowledgements}
I would like to thank Grzegorz Sitarski for useful remarks which helped to improved 
the paper. I am also very grateful to reviewers: Imre Toth and Jacek Leliwa - Kopysty\'{n}ski for valuable comments and discussions. This paper was supported by MNII grant 1 P03D 022 27. 

\section*{References}

\bibliographystyle{elsarticle-harv}

\begin{thebibliography}{99}




\bibitem[Bottke \& Melosh, 1996]{key-17}Bottke, W. F. Jr., Melosh,
H. J., 1996, {}``Binary Asteroids and the Formation of Doublet Craters'',
Icarus, 124, 372-391.

\bibitem[Bottke et al., 1999]{key-brml}Bottke, W. F., Richardson, D. C., Michel, P., Love, S. G., 1999, "1620 Geographos and 433 Eros: shaped by planetary tides ?", Astron. J., 117, 1921-1928.

\bibitem[Bottke et al., 2006]{key-bottkevar}Bottke, W. F., Vokrouhlicky D., Rubincam D. P., Nesvorny, D., 2006, "The Yarkovsky and YORP Effects: Implications for asteroid dynamics", Ann. Rev. Earth Planet. Sci. 34, 157-191.



\bibitem[Chauvineau \& Farinella, 1995]{key-cha}Chauvineau, B., Farinella, P., 1995, "The evolution of Earth-approaching binary asteroids: a Monte-Carlo dynamical model", Icarus, 115, 36-46.

\bibitem[Durda, 1996]{key-23}Durda, D. D., 1996, {}``The formation
of asteroidal satellites in catastrophic collisions'', Icarus, 120,
212-219.

\bibitem[Farinella et al., 1981]{key-13}Farinella, P., Paolicchi,
P., Zappala, V., 1981, {}``Analysis of the spin rate distribution
of asteroids'', Astron. Astroph., 104, 159-165.



\bibitem[Grundya et al., 2007]{key-grundy}Grundya, W. M., Stansberry, J. A., Noll K. S., Stephens, D. C., Trilling D. E., Kern S. D., Spencer J. R., Cruikshank D. P., Levison H. F., 2007, {}``The orbit, mass, size, albedo, and density of (65489) Ceto/Phorcys: A tidally-evolved binary Centaur'', Icarus, 191, 286-297.

\bibitem[Hadjifotinou \& Gousidou-Koutita, 1998]{key-9}Hadjifotinou, K. G., Gousidou-Koutita,
M., 1998, {}``Comparision of numerical methods for the integration
of natural satellite systems'', CM\&DA, 70, 99-113.

\bibitem[Harris, 2002]{key-harris}Harris, A. W., 2002, "On the Slow Rotation of Asteroids", Icarus, 156, 184-190.




\bibitem[Kryszczy\'{n}ska et al., 2003]{key-mich}Kryszczy\'{n}ska, A., Kwiatkowski, T., Micha{\l}owski, T., 2003,
{}``Puzzling rotation of asteroid 288 Glauke'', Astron. Astrophys., 404, 729-733.

\bibitem[Press et al., 1992]{key-11}Press, W. H., Teukolsky, S. A.,
Vetterling, W. T., Flannery, B. P., 1992, {}``Numerical recipes in
C'', second edition, Cambridge University Press.

\bibitem[Richardson et al., 1998]{key-3} Richardson, D.C., Bottke
Jr., W.F., Love, S.G., 1998, {}``Tidal distortion and disruption
of Earth-crossing asteroids'', Icarus, 134, 47-76.


\bibitem[Sharma et al., 2006]{key-19}Sharma, I., Jenkins, J. T.,
Burns, J. A., 2006, {}``Tidal encounters of ellipsoidal granular
asteroids with planets'', Icarus, 183, 312-330.

\bibitem[Scheers et al., 2004]{key-18}Scheers, D. J., Marzani, F.,
Rossi, A., 2004, {}``Evolution of NEO rotation rates due to close
encounters with Earth and Venus'', Icarus, 170, 312-323.


\bibitem[Seidelmann et al., 2007]{key-std}Seidelmann, P. K., Archinal, B. A., A’hearn, M. F., Conrad, A., Consolmagno, G. J., Hestroffer, D., Hilton, J. L., Krasinsky, G. A., Neumann, G., Oberst, J., Stooke, P., Tedesco, E. F., Tholen, D. J., Thomas, P. C., Williams, I. P.,2007, "Report of the IAU/IAGWorking Group on cartographic coordinates and rotational elements: 2006", CM\&DA, 90, 155-180.

\bibitem[Sekanina, 1982]{key-2} Sekanina, Z., 1982,{}``The problem
of split comets in review'', in Comets (L.L.Wilkening ed.), The University
of Arizona Press, Tucsun, 251-287.

\bibitem[Sekanina et al., 1994]{key-27}Sekanina, Z., Chodas, P. W.,
Yeomans, D. K., 1994, {}``Tidal disruption and the appearance of
periodic comet Shoemaker-Levy 9'', Astron. Astrophys., 289, 607-636. 

\bibitem[Sekanina et al., 1998]{key-12}Sekanina, Z., Chodas, P. W.,
Yeomans, D. K., 1998, {}``Secondary fragmentations of comet Shoemaker-Levy
9 and the ramifications for the progenitor's breakup in July 1992'',
Planet. Space Sci., 46, 21-45.

\bibitem[Sekanina \& Yeomans, 1985]{key-26}Sekanina,Z., Yeomans,
D. K., 1985, {}``Orbital motion, nucleus precession and splitting
of periodic comet Brooks 2'', Astron. J., 90, 2335-2352.

\bibitem[Sitarski, 1979]{key-10}Sitarski, G., 1979, {}``Recurrent
power series integration of the equations of comet's motion'', Acta
Astron., 29, 401-411.

\bibitem[Stansberry et al., 2008]{key-cent} Stansberry, J., Grundy, W., Brown, M., Cruikshank, D., Spencer, J., Trilling, D., Margot, J.-L., 2008, {}``Physical Properties of Kuiper Belt and Centaur Objects: Constraints from the Spitzer Space Telescope'', The Solar System Beyond Neptune, M. A. Barucci, H. Boehnhardt, D. P. Cruikshank, and A. Morbidelli (eds.), University of Arizona Press, Tucson, 592, 161-179.

\bibitem[Walsh \& Richardson, 2006]{key-20}Walsh, K. J., Richardson,
D. C., 2006, {}``Binary near-Earth asteroid formation: Rubble pile
model of tidal disruptions'', Icarus, 180, 201-216.

\bibitem[Weidenschilling, 1989]{key-15}Weidenschilling, S.
J., Paolicchi, P., Zappala, V., 1989, {}``Do asteroids have satellites
?'', in Asteroids II, M. S. Matthews, R. P. Binzel, T. Gehrels eds., University Arizona Press, Tucson, Arizona, 643-658.

\end{thebibliography}

\end{document}